\documentstyle[12pt,aasms4]{article}
\begin{document}
\begin{center}

\title{\large \bf A Calibration Method for Wide Field Multicolor \\
Photometric Systems \altaffilmark{1} }
\end{center}
\author{ Xu Zhou,Jiansheng Chen, Wen Xu, Mei Zhang}
\altaffiltext{1}{The work is supported partly by the National 
Sciences Foundation under the contract No.19833020 and No.19503003}
\affil{Beijing Astronomical Observatory,Chinese Academy of Sciences, 
Beijing 100080, China\\
 Beijing Astrophysics Center (BAC) \altaffilmark{2}, 
Beijing 100871, china 
Chinese Academy of Sciences}
\altaffiltext{2}{BAC is jointly sponsored by Chinese Academy of 
Sciences and Beijing University.}
\begingroup
\parindent=1cm

\begin{center}
Electronic mail: zhouxu@vega.bac.pku.edu.cn
\end{center}
\endgroup

\begin{abstract}
     
The purpose of this paper is to present a method to self-calibrate the 
spectral energy distribution (SED)
of objects in a survey based on the fitting of an 
SED library to the observed  
multi-color photometry. We adopt for illustrative purposes
the Vilnius (Strizyz and Sviderskiene 1972) and 
Gunn \& Stryker (1983) SED libraries. 
The self-calibration technique can improve 
the quality of observations which are not
taken under perfectly  photometric conditions. 
The more passbands used for the 
photometry, the better the results. 
This technique has been applied to 
the BATC 15-passband CCD survey. 
\end{abstract}


\section{Introduction}
Multi-color photometry can provide accurate SED information of low 
spectral resolution for a large sample of objects 
and is a very powerful tool
to deal with many important astrophysical problems. It can be used to 
measure the photometric red-shift of quasars and galaxies, and to select 
quasars and other interesting objects based on their 
characterstic SEDs. 
One can also use SEDs for stellar population synthesis, an important tool
for studying the structure and evolution of galaxies. 
The more passbands in which one observes, the better one
can determine the SED.

The Beijing-Arizona-Taipei-Connecticut (BATC) multicolor photometric survey 
is designed to obtain as much information on
the SED of celestial sources possible. 
It combines
the Beijing 
Astronomical Observatory's 60/90 cm f/3 Schmidt telescope with 
a Ford Aerospace CCD of 2048$\times$2048 pixels (recently 
replaced by a thinned Loral CCD, thanks to
the Steward Observatory) and 15 intermediate-band filters ranging from 
320 nm to 1000 nm to obtain the SEDs for all objects down to $B = 21$ 
mag in a field of one square degree. 
A key problem in the survey is calibrate accurately
the SED of our objects. 
The standard procedure of SED calibration is as follows:
on photometric nights, we observe through the 15 BATC 
filters both 
Oke \& Gunn (1983) spectroscopic standard stars 
(HD19445, HD84937, BD+26$^\circ$2606 and BD+17$^\circ$4708), 
and the target fields. 
We convolve the known fluxes of the standard stars 
with the transmission curves of the BATC filters to determine 
magnitudes in BATC photometric system (Fan et al. 1996, Zheng al. 1999). 
In practice, 
however, this method is not very efficient, for two reasons.
First, the calibrations 
are not always taken on perfectly photometric nights. 
This results in systematic troughs or bumps in the SED for all objects 
in a field.
Second, in order to obtain accurate SED calibration for 15 passbands,
we need many photometric nights, 
which could take a very long time at our 
present site. 
In order to solve these problems, we are developing 
a technique we call ``SED self-calibration''. 
This is a statistical method, 
based on applying a stellar SED library to a large sample of 
observed stellar objects. 
In section 2, we describe our method of SED self-calibration. 
In section 3 we present some results from tests of the method. 
We discussions the method and give our conclusions in section 4.

\section{Method}
The BATC survey covers regions at high galactic latitudes. 
In a typical field,  
several thousand objects are detected 
per square degree. 
Among them are several hundred  
unsaturated, bright stars with high signal-to-noise ratios.
These ``good'' stars of highly precise photometry 
form a big sample 
which is the observational basis for the SED auto-calibration method.  
In order to make the description concise, we will use, in the following 
presentation, the term $SED_{obs}$ to express the SED of 
an object which is 
calibrated by observations,
and the term $SED_{match}$ to express the SED of 
the same object which is the closest match found in the SED library.
We assume that most of these ``good'' stars are normal, nearby 
stars;
therefore, most of these ``good'' stars should have 
very close matches in the SED libraries.
If the field has been well 
calibrated by observations, the RMS of the 
differences between the 
$SED_{obs}$ and the $SED_{match}$ for most of the ``good'' stars 
should be roughly the size of the photometric precision.  
If the RMS of the residuals
is considerably larger than the photometric precision, 
then either some of 
the ``good'' stars  might not actually be normal stars,
or observations in some passbands
might not have been made during truly photometric nights.
In the former case, we can reject those stars 
with abnormal residuals from the data sample and repeat the process. 
In the 
later case, we can shift the zero points for some passbands which show 
systematic deviations of $SED_{obs}$ from $SED_{match}$. 
We iterate these processes
until the RMS of the residual reaches the
level of the photometric precision, 
which is about 0.01 mag for ``good'' stars. 
This self-calibration method is based on the following assumptions:

1. Most bright stellar objects in the field are normal stars whose SEDs can 
be found in the SED library.

2. The interstellar extinction of these bright stars is small and the 
differences in  extinction  between stars in the same CCD field 
is negligible.

3. The SED library is reliable and covers all the 
required  types of spectral and luminosity classes to fit 
the $SED_{obs}$.

We now describe briefly details of the observed and
model spectra energy distributions.

\subsection{$SED_{obs}$}
First, we must obtain $SED_{obs}$ for all the objects in the target 
field. 
Once a target field has been observed in several passbands, we 
can obtain the instrumental  colors of the stars in the field by 
using standard photometry packages, such as DAOPHOT. 
The zero point for each 
passband is obtained through observations of Oke \& Gunn
standards during photometric nights. 
As long as the zero points are well-determined, 
the instrumental 
magnitudes can be transferred into the calibrated magnitudes 
of the BATC system. 
The instrumental color index can be defined as in Table 1, where  
$Co_{j,s}^i$ is the instrument color index of j'th band minus s'th band of 
the i'th object. Here we use the $s$'th band as the reference band

\begin{table}[ht]
\center
\begin{tabular}{|l|cccc|} \hline
    -       & {\bf f1}     & {\bf f2}     & {\bf ...} & {\bf fm}     \\ \hline
   {\bf s1} & Co$^{1}_{1,s}$ & Co$^{1}_{2,s}$ &   ...     & Co$^{1}_{m,s}$ \\
   {\bf s2} & Co$^{2}_{1,s}$ & Co$^{2}_{2,s}$ &   ...     & Co$^{2}_{m,s}$ \\
   ...      &      ...     &     ...      &   ...     &      ...     \\
   {\bf sn} & Co$^{n}_{1,s}$ & Co$^{1}_{2,s}$ & 
  ...     & Co$^{1}_{m,s}$ \\\hline 
\end{tabular}
\caption{Table of the instrumental color index}
\end{table}

To calculate the flux-calibrated color index $C_{j,s}^i$, 
the color index zero 
point correction $Cc_{j,s}$ is added to the instrumental color index
$C_{j,s}$.

\begin{center}
$C_{j,s}^i=Co_{j,s}^i+Cc_{j,s}$
\end{center}

\subsection{SED libraries}
The stellar SED library used for this method is a hybrid one, including 
the theoretical SEDs of Kurucz (1992, 1993), and the observational SEDs 
of Gunn and Stryker, and of Straizyz and Sviderskiene (henceforth
referred to as ``Vilnius'').
We transfer spectral fluxes into the BATC photometric system 
using the following equation: 

\begin{center}
$	m_i  =  -2.5 log (\hat{f_{\nu}})_i -48.6 $
\end{center}

where $(f_{\nu})_i$ is the monochromatic energy at the central wavelength 
of $i^{th}$ filter, in unit of $erg cm^{-2} s^{-1} Hz^{-1}$ (cf. Fan 1995; 
Fan et al. 1996). The Vilnius library  
has 49 spectra covering spectral type from O to M6 and luminosity from main 
sequence to giant. The library of Gunn and Stryker contains 74 
spectra with 
the same coverage of the spectral type and luminosity.  The SED libraries
given by Kurucz cover temperatures from 3500K to 50000K 
with a range of 
stellar surface gravities and metallicities. 
Using these SEDs we can build 
a table of model color indices: 

\begin{table}[ht]
\center
\begin{tabular}{|l|cccc|} \hline

    -       &  {\bf f1}    &  {\bf f2}    & {\bf ...} & {\bf fm}     \\ \hline
   {\bf s1} & Cm$^{1}_{1,s}$ & Cm$^{1}_{2,s}$ &   ...     & Cm$^{1}_{m,s}$ \\
   {\bf s2} & Cm$^{2}_{1,s}$ & Cm$^{2}_{2,s}$ &   ...     & Cm$^{2}_{m,s}$ \\
   ...      &      ...     &     ...      &   ...     &      ...     \\
   {\bf sn} & Cm$^{n}_{1,s}$ & Cm$^{1}_{2,s}$ &   ...     
	& Cm$^{1}_{m,s}$ \\ \hline
\end{tabular}
\caption{Table of the model SED, where Cm$^{i}_{j,s}$ is the color index 
between the BATC j'th and the s'th filter bands of the SED library for the  
i'th spectral type and luminosity}
\end{table}

If the $SED_{obs}$ is well determined by the observations, 
it can be matched 
to one of the model SEDs with a residual at the level of photometric 
precision. 
Mathematically, we search for a minimum of $\sigma$: 

\begin{center}
$\sigma =\sum\limits_{i=1}^n\min\left[
\sum\limits_{j=1}^m\left( Co_{j,s}^i+Cc_{j,s}-Cm_{j,s}^k\right) ^2
,k \in 1,N\right] $
\end{center}

Here, $n$ is the number of the ``good'' stars, 
$m$ is the number of the color index, 
and  $N$ is the total number of entries in the model SED library. 
$Co_{j,s}^i, Cc_{j,s}$ and $Cm_{j,s}^k$ are the instrumental color index, 
the color index zero-point offset, and the color index of model SED,
respectively. 
We look simultaneously for the nearest match to each star 
in the library, and a value for
$Cc$ which leads to a minimum in $\sigma$. 
In our algorithm, 
we call subroutines of the MINUIT package (James,1994)
from the CERNLIB software.
The process converges on the 
The correct values of the color index 
zero-point corrections $Cc_{j,s}$.
During the iteration process, a 
different weight is given to each star according to its 
instrumental magnitude. 
The program can subsequently reject those stars of having abnormally high 
differences between $SED_{obs}$ and $SED_{match}$. 
At the start of the 
iteration, we must provide initial values for
the color index corrections.
The effect of different initial values on the final converged result is 
less than 0.01 mag. We take the  mean instrumental color index and the 
mean model color index as the initial corrections:

\begin{center}
$Cc_{j,s}^0=(\sum\limits_{i=1}^nCo_{j,s}^i)/n-(\sum%
\limits_{k=1}^NCm_{j,s}^k)/N$
\end{center}

Because most nearby stars are spectral type F, G and K, we employ 
only these types of model SEDs to estimate the initial color correction 
constants.

\section{Testing}
\subsection {Empirical comparison of the two methods}

We can test the method as follows:
if we observe one of the 
Oke \& Gunn standard star fields, 
we acquire the instrumental SED 
for the standard star. 
Since we know the real SED of the standard star, 
we can determine exactly the SED corrections for each passband. 
In other words, for this particular field,
the 
zero-point color corrections can be derived directly.
If we use our method on the same data set, 
we can compare its values for the zero-point corrections 
to the correct ones.

We observed the field of HD84937 through 13 filters on 
Jan. 22, 1998. 
The transparency was good, but it was not photometric. 
Two 
exposures were taken for each filter: a short one of a few seconds to 
avoid saturating the bright star HD84937,
and a long one of 300 seconds.
The short and the long exposures 
for each filter were taken in quick succession 
in order to guarantee that both shared the same weather conditions. 
The short exposure was used to determine the SED corrections for 
the field via HD84937 directly,
and the long exposure was used to determine the SED corrections
via our method.
In the following, 
we use this data set to do several tests of our 
method.

As mentioned above, we obtained the instrumental SED of standard star 
HD83927 
($MAG\_std\_instr$) from the short exposure images. Using the 
known SED of 
HD83927 in BATC system ($MAG\_std\_BATC$), we calculated the differences
in each passband $i$:

\begin{center}
   $dMAG\_std_{i} = MAG\_std\_BATC_{i} - MAG\_std\_instr_{i} $
\end{center}

If the SED self-calibration method is successful, 
it should yield corrections
for the 
SED of the long exposure images $Cc_{i}$
which differ from $dMAG\_std_{i}$ only 
by a constant $K$, due to the difference in exposure time:

\begin{center}
	$K = Cc_{i} - dMAG\_std_{i} $
\end{center}

The results are listed in Table 3. It shows that the 
constant $K$ is indeed the same in each passband,
except for the BATC2 filter. The RMS of the variation 
of $K$ in color is of the level of $0.004$. 
We give two reasons for the 
large difference in the near-UV band BATC2:
first, our thick CCD has very low quantum efficiency in 
the blue, causing low signal-to-noise values in all stars. 
Second, the SED templet in this 
region of wavelengths is available only for some stellar types. 

\begin{table}[ht]
\center
\begin{tabular}{|c|c|c|c|c|c|c|} \hline

   $Filter$ & $\lambda$ & $MAG$ &  $MAG$ 
& $Cc$ & $K$ & $K-1.14$ \\ 
               &      & $std\_instr$ &  $std\_BATC$ & & &\\ \hline
  BATC2  & 3890 & 14.475 & 8.800 & -4.3842 &  1.2908 & (+0.1508) \\  
  BATC3  & 4210 & 12.196 & 8.626 & -2.4336 &  1.1364 & -0.0036 \\
  BATC5  & 4920 & 10.342 & 8.429 & -0.8022 &  1.1107 & -0.0293 \\
  BTAC6  & 5270 & 10.118 & 8.338 & -0.6453 &  1.1346 & -0.0053 \\
  BATC7  & 5795 & 10.329 & 8.259 & -0.9308 &  1.1391 & -0.0008 \\
  BATC8  & 6075 & 10.089 & 8.232 & -0.7147 &  1.1422 & +0.0023 \\
  BATC9  & 6660 &  9.321 & 8.205 & +0.0000 &  1.1160 & -0.0240 \\
  BATC10 & 7050 & 10.500 & 8.171 & -1.1891 &  1.1399 & -0.0001 \\
  BATC11 & 7490 & 10.756 & 8.165 & -1.4294 &  1.1616 & +0.0216 \\
  BATC12 & 8020 & 10.591 & 8.150 & -1.2792 &  1.1618 & +0.0218 \\
  BATC13 & 8480 & 11.685 & 8.144 & -2.4008 &  1.1402 & +0.0002 \\
  BATC14 & 9190 & 12.317 & 8.149 & -3.0111 &  1.1569 & +0.0169 \\
  BATC15 & 9745 & 13.087 & 8.173 & -3.7746 &  1.1394 & -0.0006 \\ \hline
\end{tabular}

\caption{Results of a test of the SED self-calibration method,
using long and short exposures for the field of standard star HD84937 } 
\end{table}
		    
\subsection{Cross checking between two spectral libraries}

We can test our method in a second way.
Suppose the SED libraries, either Gunn \& Stryker or Vilnius, 
represent well the SED
for most types of stars.
Having computed synthetically 
the BATC magnitudes in all passbands for the
stars in each library,
we can apply our method to calculate the
corrections between the two sets of spectral models.
If: 

1) both libraries provide accurate stellar SEDs, and

2) each library covers an equal range of stellar types,
    i.e. any SED in one library has a corresponding SED in 
         the other library, and 

3) the method developed here is correct and effective, 

then
after the iteration 
process, 
the final SED corrections should be close to zero.

In the following test, we use Gunn \& Stryker library as $SED_{match}$ 
and the  Vilnius library as the $SED_{obs}$ catalog. 
The results are shown in 
Table 4. 
Column 1 is the filter name, 
column 2 its central wavelength,
Column 3 the correction for each band. 
The BATC9 passband (6660A) 
is used as the reference band, and kept fixed during the iterative process. 
The final 
values for the zero point corrections are indeed very small. 
The mean 
deviation of zero point corrections is $\pm0.015$ mag;
however, the BATC1 band again shows much larger devaitions 
than the other bands. 

This result shows that the two SED libraries in general can be matched well 
each other, but two points call for for further discussion. 

1) There is a systematic deviation for both SED libraries 
in the blue versus the red:
below 455nm, the Vilnius SEDs are flatter than those of Gunn \& Stryker;
above 455nm, the Vilnius SEDs are more depressed those of Gunn \& Stryker.

2) The large deviation in UV band indicates that either one or both 
the SEDs does not well represent the stellar SEDs in this region.

\begin{table}[ht]
\center
\begin{tabular}{|c|c|cc|} \hline
Filter & $\lambda$ &  model       &  note  \\
       &           &  calibration &        \\ \hline
BATC1  &   3360    &	  -0.041      &        \\
BATC2  &   3890    &  -0.019      &        \\       
BATC3  &   4210    &  -0.029      &        \\       
BATC4  &   4550    &  -0.009      &        \\       
BATC5  &   4920    &   0.003      &        \\       
BATC6  &   5270    &   0.025      &        \\       
BATC7  &   5795    &   0.010      &        \\       
BATC8  &   6075    &   0.008      &        \\       
BATC9  &   6660    &   0.000      &  fixed \\       
BATC10 &   7050    &   0.005      &        \\       
BATC11 &   7490    &   0.018      &        \\       
BATC12 &   8020    &   0.008      &        \\       
BATC13 &   8480    &   0.008      &        \\       
BATC14 &   9190    &   0.010      &        \\       
BATC15 &   9745    &   0.007      &        \\ \hline
\end{tabular}
\caption{Results of cross checking between Vilnius 
            and Gunn \& Stryker model spectra}
\end{table}

\begin{figure}[tt]
\plotone{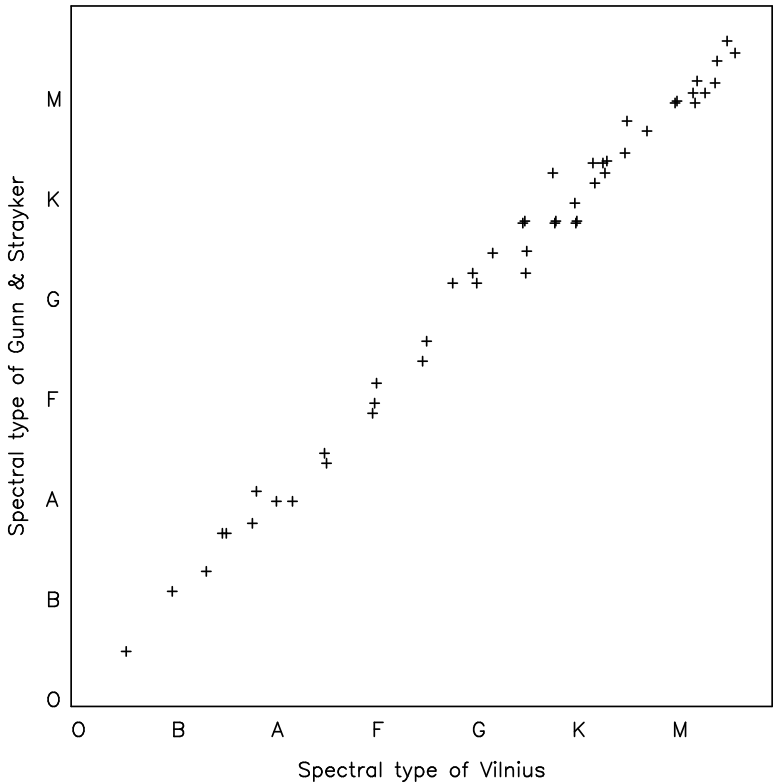}
\caption{Spectral type of Vilnius vs spectral 
types obtained from Gunn \& Stryker model}
\end{figure}

\subsection{Stellar classification}
The process of matching $SED_{obs}$ to $SED_{match}$ 
produces as a byproduct
a spectral classification for each star;
or, if a library of theoretical models is used,
the stellar atmospheric parameters of each star. 
It provides us 
with an indirect way to test the method presented in this paper. 
Figure 1 
shows a good correlation between 
the original spectral class of stars in the Vilnius catalog,
and the spectral class determined for those stars via our method,
using the Gunn \& Stryker library for $SED_{match}$.

\section{Discussions and conclusions}

1. Our method is a statistical one, 
so it can only be effectively applied 
to a large stellar sample. The BATC survey satisfies this condition. 
Each BATC field covers about one square degree. 
Typically, there are more 
than 4000 objects detected in each image. 
We are guaranteed 
several hundred bright and unsaturated stars 
with reliable instrumental magnitudes 
to use as a ``good'' star sample.

2. The basis of our method is fitting the observed stellar SED 
to a library of stellar SEDs. 
Our assumption that most of our 
``good'' stars are normal stars appears firm. 
and their SEDs can 
be found from the 
SED library. 
Most of abnormal stars are rejected 
during of the iteration process.

3.  The SED library is a key for the method. 
There is a great demand 
in the astronomical community for reliable SED libraries 
which cover a wide range in wavelength and stellar type,
The theoretical libraries 
(e.g. Kurucz 1993) are limited by our knowledge of stellar physics. 
The theoretical SEDs for late type, low-temperature stars is
not as reliable as those for hot stars. 
There are many observational stellar 
libraries (Gunn \& Stryker, Vilnius, etc),
but differences among them exist. 
It is not easy to judge which one is the best. 
After comparing various 
stellar SED libraries we have collected from the literature, 
we are at present using mostly
the Gunn \& Stryker library in our calculations. 
Further work on making good libraries of SEDs is necessary.

4. In principle, interstellar extinction must be considered in the final 
result. Our method of 
SED correction takes into account not only terrestrial atmospheric 
extinction, but also interstellar reddening. 
If standard stars or spectral models
suffer from systematically more or 
less (or different) extinction than a survey's target stars,
our method will yield systematically incorrect zero-point corrections
to the photometry of target stars.
This is because the method 
uses a fitting process to fit the observational SED affected by the 
reddening to the theoretical SED unaffected by the reddening.  
In the BATC 
survey, the fields are located at high galactic latitudes, where the 
interstellar extinction is much reduced. 
We therefore 
use only bright stars, which are nearby and little affected by 
interstellar extinction, as standards for the survey fields.

5.  We have made several tests to see if the method works.

(1) We have tried different filter bands as the reference band 
to see if the change 
in the reference band affects the results. Our results
show that the difference in 
the final constant corrections is less than 0.01 mag. This means that the 
choice of reference band in color index is not important. 
Normally, we select the band of deepest exposure as the reference band, 
since it has highest signal to noise ratio. This also allows the
measurement of color index for as many stars as possible.

(2) We wanted to know if the number of the filters used affects the results,
and what is the minimum number of colors required to 
make the SED self-calibration work effectively. 
In order to do the test, we reduced the number of filters by 
taking off some of the color indexes. 
We found that the method still
yields reasonable results. Obviously, the results improve when 
one uses a larger number of the filter bands, 
and a wider range of wavelengths.

(3). We wanted to know if the final results depend on the characteristics of 
the ``good'' stars, which we used for the SED self-calibration. For this 
purpose, we used various randomly selected subsets of the ``good'' stars 
in each image.
We obtained very similar color correction constants. 
Furthermore, we divided the ``good'' stars sample into several sub-groups 
according to their apparent magnitude. 
The results show that the difference 
among different groups is about 0.03 mag.  
One possible reason is the low signal-to-noise ratio of faint stars.

(4) Does the choice of the initial value of the 
iteration affect the results? 
Our tests show that different initial values normally only affect the 
convergence time of iteration. The difference on the final result is less 
than 0.01 mag.  In some cases, there are several minima and the 
process may not iterate to the right one; this is an open issue.
The best way to solve this problem, from our experiences, is that if one or 
a few color indexes are very well determined by observations taken 
during photometric nights, then we keep these
indexes fixed. This causes the iteration always to converge
to the right minimum.

  After many tests and applications to real data, we conclude that, 
though there are still some problems requiring further development
(such as creating larger libraries of stellar SEDs),
the method presented here can work well:
the accuracy of the SED calibration 
is comparable to the precision of the CCD photometry. 
A by-product of our method is the automatic classification of the 
stellar type or the determination of the stellar parameters, which is 
very useful for the studies of galactic structure 
via large field multi-color CCD photometry survey. 

\acknowledgments

\end{document}